\documentclass{IEEEtran}
\usepackage{cite}
\usepackage{amsmath,amssymb,amsfonts}
\usepackage{graphicx}
\usepackage{textcomp,nicefrac}
\def\BibTeX{{\rm B\kern-.05em{\sc i\kern-.025em b}\kern-.08em
T\kern-.1667em\lower.7ex\hbox{E}\kern-.125emX}}
\begin{document}
\title{Higher Order Nyquist Zone Sampling with RFSoC Data Converters for Astronomical and High Energy Physics Readout Systems}
\author{Chao Liu, Zeeshan Ahmed, Shawn W. Henderson, Ryan Herbst and Larry Ruckman 
\thanks{
All the authors are with the SLAC National Accelerator Laboratory , Menlo Park, CA 94025, USA (e-mail: chaoliu@slac.stanford.edu).}
}

\maketitle

\begin{abstract}
From generation to generation, the maximum RF frequency and sampling rate of the integrated data converters in RF system-on-chip (RFSoC) family devices from Xilinx increases significantly. With the integrated digital mixers and up and down conversion blocks in the datapaths of the data converters, those RFSoC devices offer the capability for implementing a full readout system of ground and space-based telescopes and detectors across the electromagnetic spectrum within the devices with minimum or no analog mixing circuit. In this paper, we present the characterization results for the the data converters sampling at higher orders of Nyquist zones to extend the frequency range covered for our targeted readout systems of microwave-frequency resonator-based cryogenic detector and multiplexer systems and other astronomical and high-energy physics instrumentation applications, such as, axion search and dark matter detection. The initial evaluation of the data converters operating higher order Nyquist zones covers two-tones and comb of tones tests to address the concerns in the RF inter-modulation distortion, which is the key performance index for our targeted applications. The characterization of the data converters is performed in the bandwidth of 4-6 GHz and results meet our requirements. The settings and operating strategies of the data converters for our targeted applications will be summarised. 
\end{abstract}

\begin{IEEEkeywords}
Microwave, Spectrometry, Data Converter, Sampling, Readout.
\end{IEEEkeywords}

\section{Introduction}
\label{sec:introduction}
\IEEEPARstart{T}{he} majority of  past application of RFSoC leverage the developed analog RF up and down mix circuit to convert the RF signal with in the Nyquist frequency of the converters. Both other research teams \cite{chao}, \cite{chao1} and us \cite{smurf}, \cite{spacesmurf} have done comprehensive performance characterization of the integrated data converter in first order Nyquist zone and proven the signal to noise ratio, the dynamic range, effective number of bits and cha distortion (IMD) are within the specifications of our target applications with both Xilinx evaluation board and custom designed board. In this paper, we will summaries the evaluation results for data converters operate in higher order Nyquist zones. Since IMD level is one of the most critical index for the highly multiplexed readout, the paper will focus on the discussion of configurations of data converter have significant impact in the IMD level.

\begin{figure}[t]
\centerline{\includegraphics[width=3.5in]{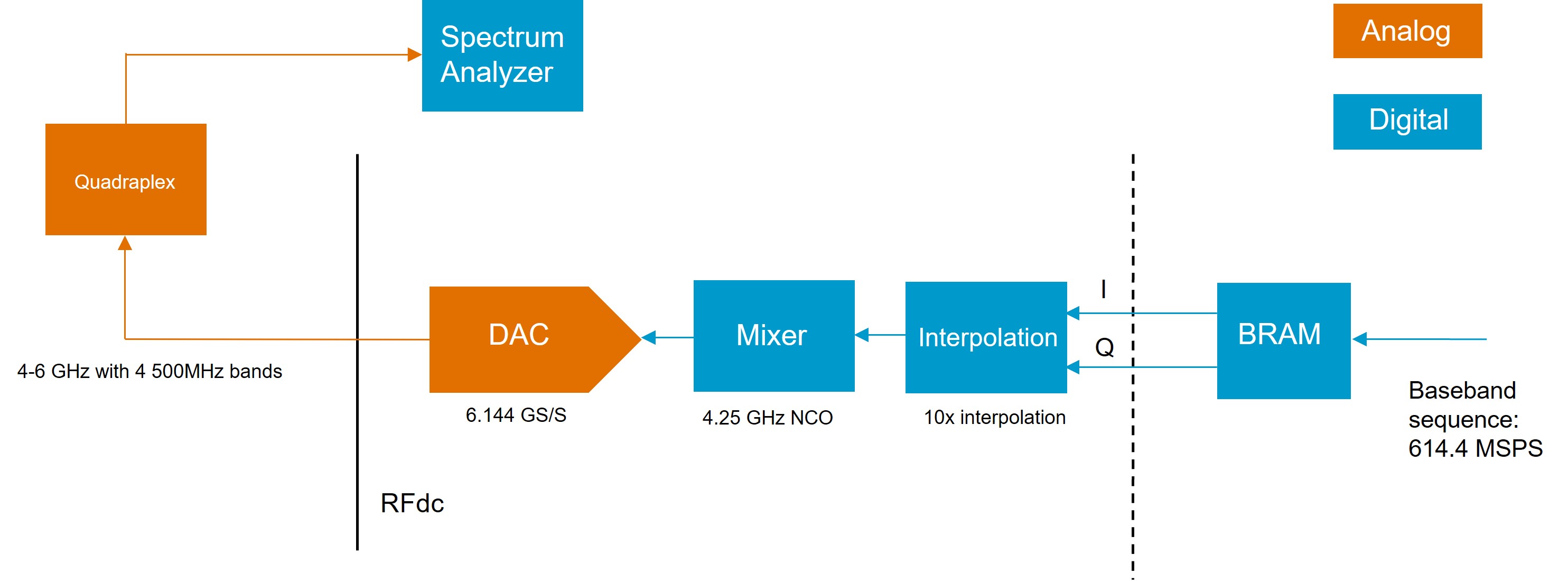}}
\caption{DAC performance evaluation setup.}
\label{fig1}
\end{figure}

\begin{figure}[t]
\centerline{\includegraphics[width=2.5in]{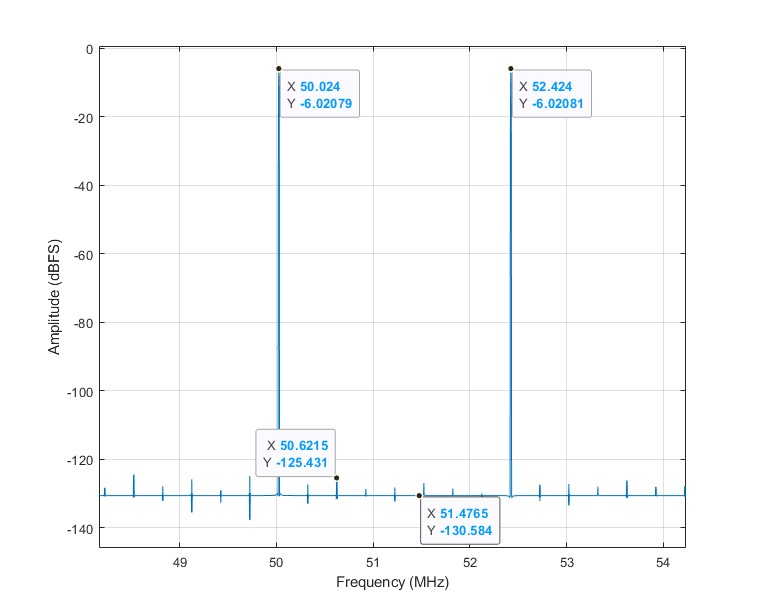}}
\caption{FFT of two-tone sequence loaded to the DAC.}
\label{fig2}
\end{figure}

The prototype of the readout system uses the Xilinx Zynq UltraScale+ RFSoC ZCU208 Evaluation Kit, which carries a Gen RFSoc device with eight 14-bit ADCs up to 5GSPS , and eight 14-bit DACs up to 10GSPS. All the tests describe in this paper is performed with ZCU208.

\section{Data Converter Performance Characterization at Higer Order Nyquist Zones}

The first step of characterization is to evaluate the signal generated by the DAC. Figure \ref{fig1} shows the full test circuit for DAC performance measurement. In this case the baseband quadrature sequence is loaded to BRAM and streamed to the RF data converter (RFdc) at the data rate of 614.4 MHz. The quadrature sequence is interpolated by factor of 10 and up-mixed at 4.25 GHz with the integrated blocks in DAC datapath. In this case, the DAC is sampling at 6.144 GSPS and carrier frequency is in the second Nyquist zone of the DAC. The signal generated by the DAC is filtered by a bandpass filter before injected to the spectrum analyzer, which is the Keysight EXA signal analyzer N9010B with bandwidth from 10 Hz to 26.5 GHz. The IMD performance of digital up-conversion and second Nyquist zone is investigated with two-tone test and comb of tones generation for our applications. 

\subsection{Two-tone Test}

Two-tones test is the most common test to evaluate the IMD performance of radio frequency electronic system. Figure \ref{fig2} shows the FFT of the two-tone baseband sequence loaded to BRAM for testing. The first tones is around 50 MHz and the second is 2.4 MHz higher. 
\begin{figure}[t]
\centerline{\includegraphics[width=3 in]{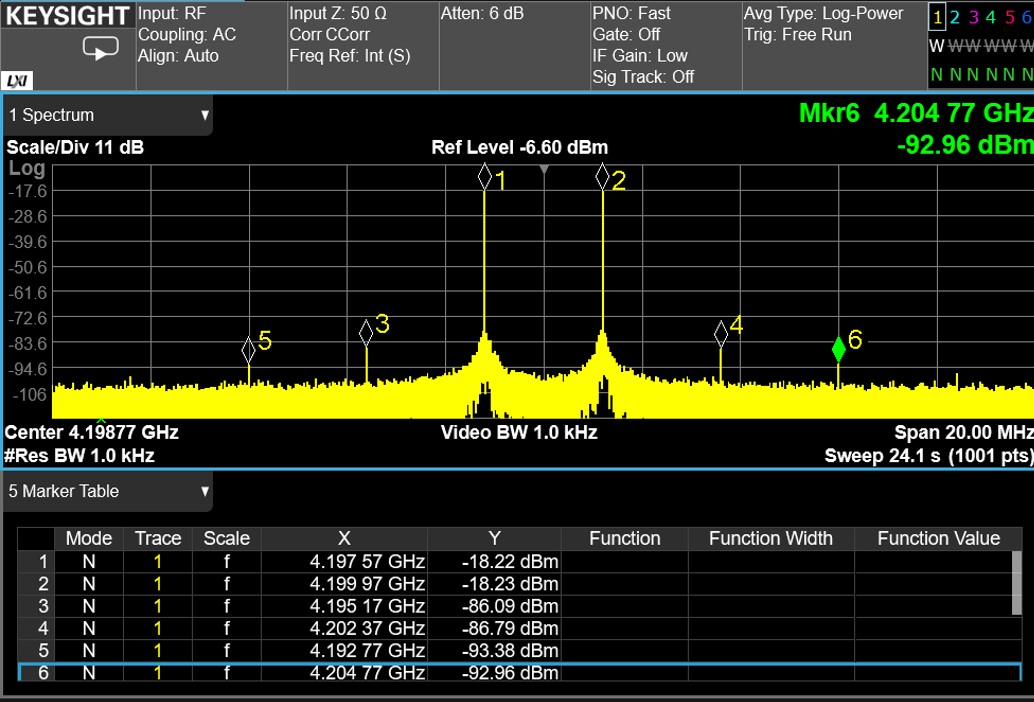}}
\caption{Spectrum of DAC output at SNR optimized decoder mode.}
\label{fig3}
\end{figure}

\begin{figure}[t]
\centerline{\includegraphics[width=3 in]{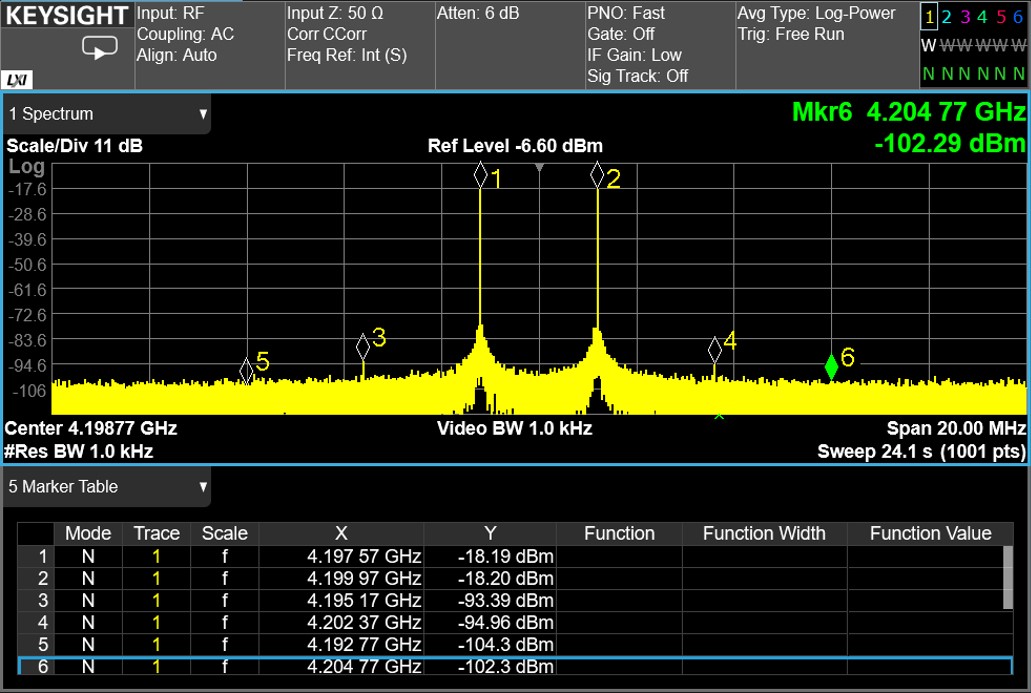}}
\caption{Spectrum of DAC output at high linearity decoder mode.}
\label{fig4}
\end{figure}
\section{DAC Performance Characterization}
As the signal generated by DAC locates in the second order Nyqusit zone, the DAC has been configured to RF mix mode. The RF mix mode has adopted the pulse type with second half inverted, which can maximize the energy in second Nyquist zone. The decoder mode of the DAC is also a critical settings for achieving lower intermodualtion level. Figure \ref{fig3} and \ref{fig4} shows the spectrum measured by the signal analyzer for DAC output in the two decoder modes, SNR optimized and high linearity. The DAC consists of unit current cells turned on and off to achieve corresponding digital code. The current cells all have systematic static DC error and the fixed selection current cell can result in higher intermodulation products. The high linearity mode randomized the selection of current cell to spread the error over the whole Nyquist band. From Figure \ref{fig3} to \ref{fig4}, the third order and higher order products are reduced by at least  7.3 dB without significant increase in the noise floor level. The high linearity mode achieved 76 dB clearance between the tones and intermodualtion products and should be used for IMD sensitive applications. The amplitude of signal and DAC current level need to be set to operate 50 \% lower than the maximum rate to achieve the optimum IMD level.  

\subsection{Comb of Tones Test}

\begin{figure}[t]
\centerline{\includegraphics[width=3 in]{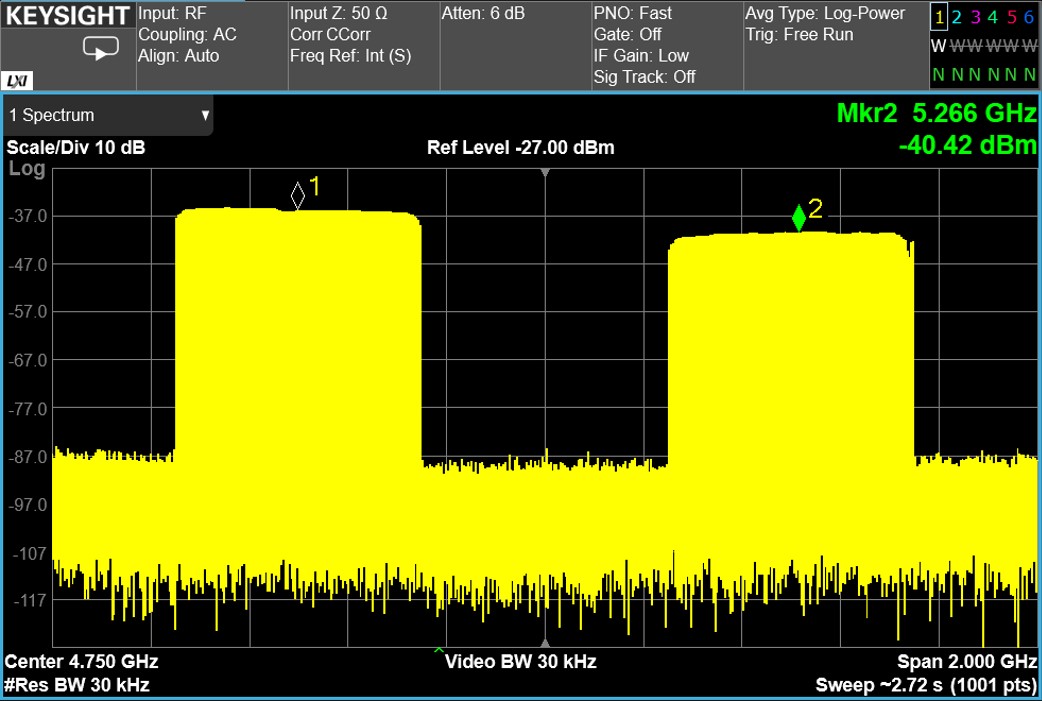}}
\caption{Spectrum of DAC output at full amplitude range.}
\label{fig5}
\end{figure}

The RFSoC based readout between 4-6 GHz needs to divided to four 500 MHz block based on previous generation readout system. In this test, two of the blocks centred at 4.25 GHz and 5.25 GHz have been generated by the DAC. There are 209 tones in steps randomized around 2.4 MHz in this test. From the spectrum capture by the signal analyzer shown in Figure \ref{fig5}, the two blocks are generated with low level of intermodulation products and leakage to other bands.

\section{Conclusions}

The IMD level of the integrated DAC datapath with appropriate configuration is optimistic to meet the stringent requirement of highly multiplexed readout. Similar two-tone characterization has also been performed for the integrated ADC datapath in third order Nyquist zone and it also gives about 75 dB clearance between the tones and the third order intermodualtion product. Therefore, the IMD level of the integrated datapaths in RFSoC has the potential be used to realize readout system between 4-6 GHz or even higher frequency without any analog RF mixing circuit.

\end{document}